# Emergence of a Metal-Insulator Transition and High Temperature Charge Density Waves in VSe$_2$ at the Monolayer Limit

Ganbat Duvjir,[†] Byoung Ki Choi,[‡] Iksu Jang,[§] Søren Ulstrup,[∥,⊥] Soonmin Kang,[#,∇] Trinh Thi Ly,[†] Sanghwa Kim,[†] Young Hwan Choi,[‡] Chris Jozwiak,[∥] Aaron Bostwick,[∥] Eli Rotenberg,[∥] Je-Geun Park,[#,∇] Raman Sankar,[∘,◆] Ki-Seok Kim,[*,§] Jungdae Kim,[*,†] Young Jun Chang,[*,‡]

[†]Department of Physics, BRL, and EHSRC, University of Ulsan, Ulsan 44610, Republic of Korea
[‡]Department of Physics, University of Seoul, Seoul 02504, Republic of Korea
[§]Department of Physics, Pohang University of Science and Technology (POSTECH), Pohang 37673, Republic of Korea
[∥]Advanced Light Source (ALS), E. O. Lawrence Berkeley National Laboratory, Berkeley, California 94720, USA
[⊥]Department of Physics and Astronomy, Interdisciplinary Nanoscience Center, Aarhus University, 8000 Aarhus C, Denmark
[#]Center for Correlated Electron Systems, Institute for Basic Science (IBS), Seoul 08826, Republic of Korea
[∇]Department of Physics and Astronomy, Seoul National University (SNU), Seoul 08826, Republic of Korea
[∘]Institute of Physics, Academia Sinica, Taipei 10617, Taiwan
[◆]Center for Condensed Matter Sciences, National Taiwan University, Taipei 10617, Taiwan

[*]e-mail: tkfkd@postech.ac.kr; kimjd@ulsan.ac.kr; yjchang@uos.ac.kr

**Abstract**

Emergent phenomena driven by electronic reconstructions in oxide heterostructures have been intensively discussed. However, the role of these phenomena in shaping the electronic properties in van der Waals heterointerfaces has hitherto not been established. By reducing the material thickness and forming a heterointerface, we find two types of charge-ordering transitions in monolayer $VSe_2$ on graphene substrates. Angle-resolved photoemission spectroscopy (ARPES) uncovers that Fermi-surface nesting becomes perfect in ML $VSe_2$. Renormalization group analysis confirms that imperfect nesting in three dimensions universally flows into perfect nesting in two dimensions. As a result, the charge density wave transition temperature is dramatically enhanced to a value of 350 K compared to the 105 K in bulk $VSe_2$. More interestingly, ARPES and scanning tunneling microscopy measurements confirm an unexpected metal-insulator transition at 135 K, driven by lattice distortions. The heterointerface plays an important role in driving this novel metal-insulator transition in the family of monolayered transition metal dichalcogenides.

**Keywords:** charge-density wave, vanadium diselenide, metal-insulator transition, transition metal dichalcogenides, two-dimensional materials.

Electronic reconstruction, caused by the surface/interface modification of correlated-electron behavior, is a fundamental scientific issue and has been extensively applied to oxide heterostructures.[1, 2] An understanding of electronic reconstruction due to dimensionality and heterointerface coupling in layered van der Waals systems remains an open issue. When reduced to the monolayer (ML) limit, layered transition metal dichalcogenides (TMDs) have shown strong dimensionality effects in numerous fundamental properties, such as band gap type,[3] superconductivity,[4] charge density wave (CDW) formation,[5] and ferromagnetism.[6] Although the influence of the interface is often considered to be negligible for the TMDs due to the weakness of the weak van der Waals interactions, the substrate may play an important role due to strain[7] and dielectric screening.[8]

Vanadium diselenide ($VSe_2$), a member of the metallic TMD family, presents a rare example of a three-dimensional (3D) CDW phase that has a ($4a \times 4a \times 3c$) nesting vector with a transition temperature ($T_{CDW}$) of 105 K and coexists with itinerant electrons of the residual Fermi surface (FS).[9, 10] The CDW distortion is small compared to the atomic corrugation, mostly due to the weak 3D nesting condition.[11] The reduced dimensionality in thin flakes and ultrathin films of $VSe_2$ has led to a highly debated thickness-dependent variation of $T_{CDW}$, which has been difficult to reproduce.[6, 12, 13] It is noted that $VSe_2$ films could possess different structural and electronic orderings, such as a moiré structure on $MoS_2$ substrates[6] and an insulating ($4 \times \sqrt{3}$) surface reconstruction on $Al_2O_3$ substrates[7], implying an important role of heterointerface for tuning electronic and structural properties of $VSe_2$. Furthermore, recent studies have reported ferromagnetic ground states in ML and few layer $VSe_2$ films on highly oriented pyrolytic graphite (HOPG) substrates.[6, 14, 15]

Here, we report systematic studies on the electronic and atomic structures of ML VSe$_2$ epitaxially grown on bilayer graphene (BLG) on silicon carbide (SiC) using both angle-resolved photoemission spectroscopy (ARPES) and scanning tunneling microscopy (STM). We observe the emergence of a metal-insulator transition (MIT) and a high-temperature CDW phase, associated with heterointerface coupling and reduced dimensionality, respectively. Temperature-dependent ARPES measurements reveal that perfect FS nesting enhances the CDW transition temperature such that $T_{CDW} = 350 \pm 8$ K. Renormalization group (RG) analysis confirms that the two-dimensional (2D) nature of the ML VSe$_2$ drives the perfect FS nesting. STM measurements confirm that the CDW order exists both at 300 K and 79 K. We also observe an unexpected MIT with a transition temperature of $T_{MIT} = 135 \pm 10$ K, driven by strong lattice distortion of Se atoms. The lattice distortion is attributed to the dimerization of V atoms which stabilizes the insulating phase.

The structure model of ML 1T-VSe$_2$ on BLG is given in Figure 1a. Figure 1b shows a STM image of a VSe$_2$ film on BLG at a coverage of 0.9 ML (see in Supporting Information (SI) Figure S1). The line profile along the arrow in Figure 1b shows that the apparent height of the VSe$_2$ film is 6.9 Å which agrees well with the unit cell height of bulk VSe$_2$ (6.1 Å).[9] Although the lattice mismatch between VSe$_2$ and graphene is quite large, about 26.5%, their crystal axes are aligned within the rotational misalignment $\theta \leq \pm 5°$[3, 4, 8] (Figure S2-S4).

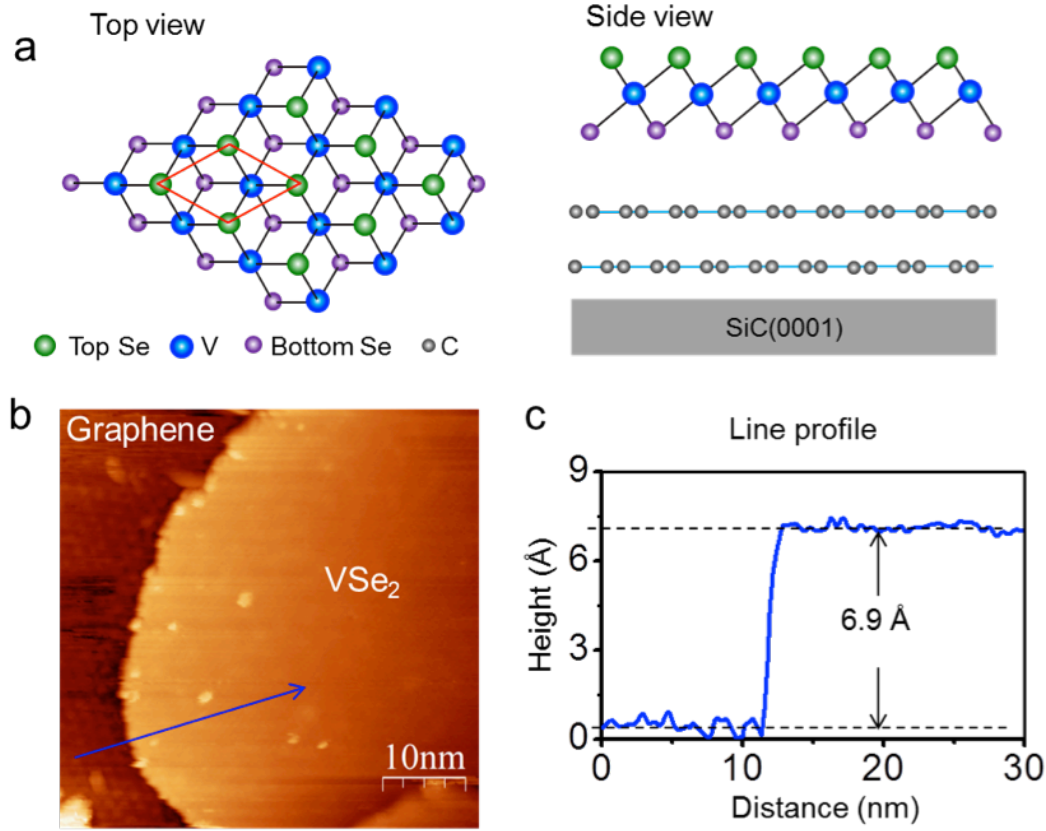

**Figure 1.** Morphology of ML VSe$_2$ on BLG. (a) Top- and side-view schematics of ML VSe$_2$ with a BLG substrate where the green, purple, blue and gray balls represent the top Se, bottom Se, V, and C atoms, respectively. (b) Topographic STM image of 0.9 ML VSe$_2$ grown on BLG ($V_b$ = -1.2 V, $I_t$ = 40 pA). (c) Line profile along the arrow in (b).

Our ARPES measurements in Figure 2a show the FS map of ML VSe$_2$ at 180 K. Six cigar-shaped contours enclose long parallel sides associated with a strong nesting condition, as outlined by the solid lines. Based on the measurement, we suggest a simplified FS model of ML VSe$_2$, showing a hexagonal Brillouin zone with six cigar-shaped electron pockets centered at the M points in Figure 2b.[15] It should be noted that the FS structure of ML VSe$_2$ qualitatively differs from that of bulk VSe$_2$ (Figure S5a-c). Bulk VSe$_2$ exhibits a 3D FS with a strong $k_z$ dispersion where the main features can be described as repetition between symmetric elliptical electron pockets and asymmetric dogbone pockets.[10] In contrast, ML VSe$_2$ possesses an FS with a 2D character, which is non-dispersive along the k$_z$ direction (Figure S5e). Figures 2c-d show ARPES spectra collected along the M-Γ-M and K-M-K directions, respectively, which has slightly broadened features due to the crystalline misalignment. In Figure 2c, the upper band 'A' crosses the Fermi level, while the maxima of the other two dispersive bands 'B' and 'C' are located below the Fermi level. Figure 2b displays parabolic band structure, where FS nesting occurs between the two Fermi level crossing points (marked with arrows).

As shown in Figure 2e-f, we monitor the temperature-dependent energy dispersion curves (EDCs) during cooling at the fixed values of k, marked in Figures 2a, 2c, and 2d, referred to as the cold spot 'α' (a residual FS segment) and hot spot 'β' (a gapped segment due to a strong CDW interaction). Above 135 K, the midpoint of the leading edge (red dashed lines) corresponds to the Fermi level (black dashed lines) at α, while the edge is shifted below the Fermi level at β, where an energy gap of 30 meV is determined. At 135 K, we observe a sudden energy shift of the spectra at both α and β, suggesting an abrupt electronic phase transition. The energy shift is reproducible during warming (Figure S5f-i).

In Figure 2g, we track the electronic gap (Δ) along the dashed line in Figure 2a, which reveals two types of electronic ordering phenomena in ML VSe$_2$. Considering the partial gap opening near β in the temperature range from 150 to 300 K and the strong FS nesting, it is suggested that the CDW phase exists above 300 K. At 135 K, the gap is fully opened for all values of k with a minimum gap size of 9 ± 4 meV at α, directly indicating an MIT with T$_{MIT}$ = 135 ± 10 K. This MIT has never been observed in bulk VSe$_2$. The gap values extracted at β are fitted with the Bardeen-Cooper-Schrieffer (BCS) formula in Figure 2h. The squared gap values exhibit a linear dependence over a wide temperature range. In particular, the $T_{CDW} = 350 \pm 8$ K estimated from the BCS formula is much higher than that of the bulk (105 K). We argue that this drastic enhancement of $T_{CDW}$ originates from the perfect FS nesting due to the long straight FS contours. Such straight FS segments are often observed in 1D materials, such as quasi-1D bulk,[16, 17] atomic chains,[18] and ultrathin oxide films.[19] Our RG analysis suggests that the emergence of such perfect FS nesting is universal, which we discuss in detail below.

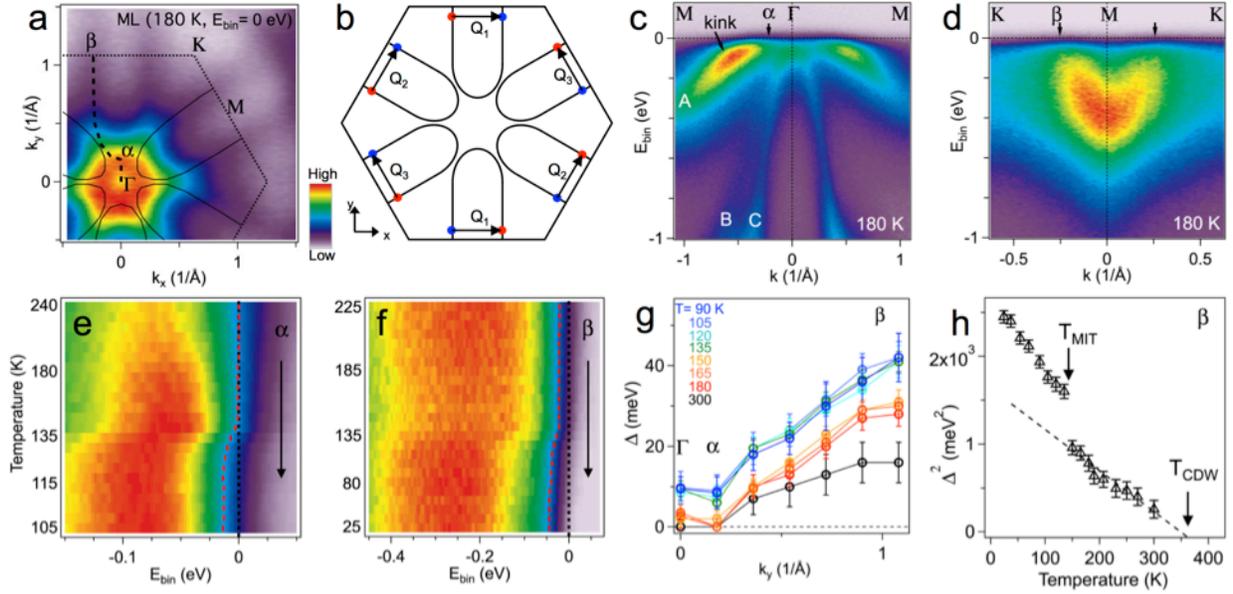

**Figure 2.** Band structure and temperature dependence of ML VSe$_2$. (a) Fermi surface map at 180 K. The solid lines outline the peak intensity of the FS contours. The dotted line indicates the first Brillouin zone of VSe$_2$. (b) Schematic Fermi surface contour of ML VSe$_2$. $Q_n$ with n = 1, 2, 3 represents a nesting wave vector of each pocket. (c) and (d) ARPES spectra along M-Γ-M and K-M-K for ML VSe$_2$ (180 K). (e) and (f) Temperature dependent energy dispersion curves at the α and β points, revealing an abrupt energy gap opening at ~135 K during cooling. Red and black dashed lines represent the midpoints of the leading edges and the Fermi level, respectively. (g) Energy gap (Δ) profile at various temperatures along the dashed line in (a) shows a partial gap opening around the β point and a complete gap opening below the MIT temperature ($T_{MIT}$ = 135

± 10 K). (h) Temperature dependence of the energy gap at the β point shows BCS-type scaling behavior for the CDW phase and a sudden discontinuity at $T_{MIT}$. The CDW transition temperature ($T_{CDW}$ = 350 ± 8 K) is estimated from extrapolation of the linear approximation (dashed line) of the data above $T_{MIT}$.

Our STM measurements reveal the atomic scale structure associated with both the room-temperature CDW metallic phase and the low-temperature insulating phase. Figures 3a and 3c show filled state STM images of ML VSe$_2$ obtained at 79 K and 300 K, respectively. At 300 K, the typical hexagonal structure of the top Se layer is observed (see Figure 3c). At a lower temperature of 79 K, the atomic structure drastically changes, giving rise to ($\sqrt{3} \times 2$) and ($\sqrt{3} \times \sqrt{7}$) superstructures as shown in Figure 3a. A detailed line profile analysis, conducted along the arrows in Figures 3a and 3c, indicates that the low-temperature phase possesses strong lattice distortions in both the vertical and lateral directions compared to the room-temperature phase, as shown in Figure 3g. These observed distortions are related to the formation of Se-Se dimers, which are laterally paired to be ~ 2.8 Å (12.5% reduced) with respect to the undistorted Se-Se distance (3.2 Å, 300 K). Such dimerization is also reflected as strong peaks (red circles) in the fast-Fourier transformation (FFT) image of Figure 3b (see Figure S6 for discussion on the additional peaks). Since the ($\sqrt{3} \times 2$) and ($\sqrt{3} \times \sqrt{7}$) superstructures disappear at 300 K, we can rule out the possibility that the observed superstructure are driven by the moiré between VSe$_2$ and graphene. This also agrees with recent STM studies of ML VSe$_2$ grown on HOPG at 150 K[6] and 300 K[20].

Figure 3(l-o) and 3(q-t) present topographic and corresponding FFT images of ML VSe$_2$ obtained from different locations of sample, respectively. Rotational misalignments of θ ≤ ±5°

exist between graphene (white dashed line) and ML VSe$_2$ (red dashed line). Although the main peaks of lattice distortion (red circles) are consistently observed in all FFT images of Fig. 3q-t, we note that detailed topographic features vary in Fig. 3l-o, which don't fit well with ($\sqrt{3}\times 2$) and ($\sqrt{3}\times\sqrt{7}$) superstructures. The dependence of superstructures on the misalignment indicates that interface coupling between graphene and ML VSe$_2$ should play an important role in this heterostructure system. From the symmetry of hexagonal lattices, stripe pattern should be allowed in 3 different orientations, and figure 3n shows two orientations adjacent to one another. Considering only one crystal axis along the dimerized direction (blue arrow in Figure 3a), the ($\sqrt{3} \times 2$) and ($\sqrt{3} \times \sqrt{7}$) superstructures match with $12 \times a_G$ (graphene lattice constant), suggesting heterointerface coupling for this system (Figure S7). However, better structural model requires sophisticated theoretical analysis of two-dimensional lattices.

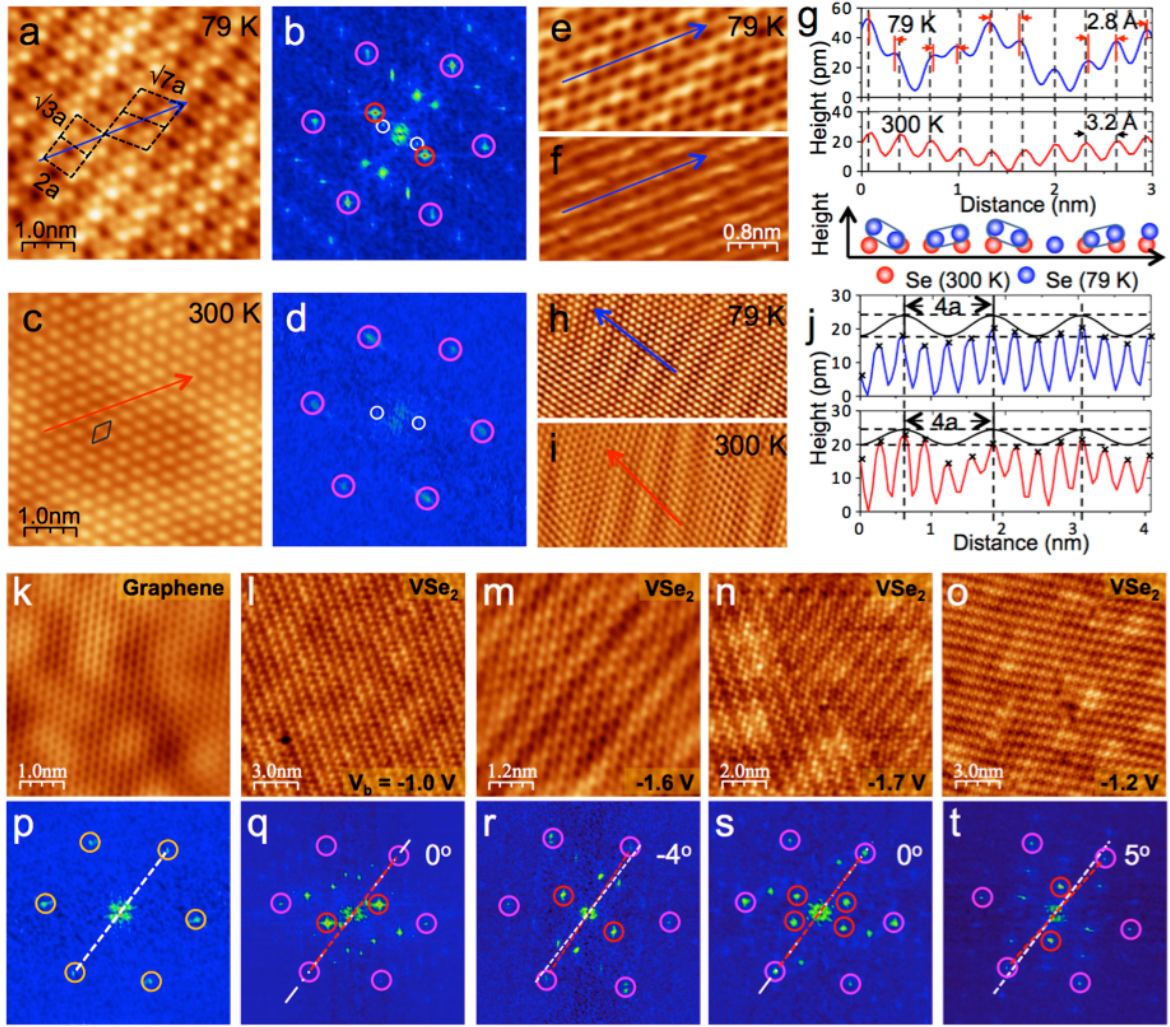

**Figure 3.** STM analysis of ML VSe$_2$. (a) and (c) STM topographic image of ML VSe$_2$ at 79 K (V$_b$= -1.4 V, I$_t$ = 30 pA) and 300 K (V$_b$= -0.8 V, I$_t$ = 100 pA), respectively. In (a) (√3 × 2) and (√3 × √7) superstructures are marked by the dotted black lines. In (c) the VSe$_2$ unit cell is indicated as a black diamond. (b) and (d) 2D FFT images of larger STM topographies (20 × 20 nm$^2$) at 79 K and 300 K, respectively. The pink, white, and red circles indicate peaks associated with Bravais lattice points, CDW structure, and the lattice distortion of VSe$_2$, respectively. (e) STM topography and (f) dI/dV maps obtained simultaneously (V$_b$= +1.6 V, I$_t$ = 30 pA, 79 K). (g) Line profiles measured along the blue arrow in (a) (79 K) and the red arrow in (c) (300 K), which track the positions of top Se atoms. The lattice distortion leads to dimerized pairs of Se atoms at 79 K, indicated with red arrows. The vertical dashed lines mark the undistorted lattice constant of 3.2 Å. The bottom inset describes the atomic arrangements of Se atoms at 300 K and 79 K. (h) and (i) Fourier-filtered images (10 × 5 nm$^2$) at 79 K and 300 K, which were obtained by selectively retaining the pink and white circles in (b) and (d). (j) The blue and red profiles are obtained along the arrows in (h) and (i). From the peaks of atomic corrugation marked by '×', the CDW modulations are visible at both 79 K (blue) and 300 K (red). The black curves indicate the sinusoidal fitting (CDW modulation) to the marked corrugation peaks. (k) and (p) Topographic and corresponding FFT images of graphene substrates, respectively. (l-o) and (q-t) Topographic and corresponding FFT images of ML VSe$_2$, obtained from different locations. The misalignment angles between graphene (white dashed line) and ML VSe$_2$ (red dashed line) are indicated in (q-t).

In order to understand the relation between the lattice distortion and the MIT, we compare our VSe$_2$ film with VO$_2$, a strongly correlated material with a first-order MIT above room temperature. The microscopic origin of the MIT in VO$_2$ is the dimerization of V$^{4+}$ ions along the edge-shared chain of VO$_6$ octahedra, where V 3$d$ electrons are localized to form the spin singlet state, and both electron-electron and electron-phonon interactions are generally considered to be important.[21, 22] Since the 1T-VSe$_2$ structure has similar edge-shared VSe$_6$ octahedra, we propose that the distorted top-Se atoms reflect the dimerization of V atoms in the insulating phase. It was reported that V 3d orbital states dominate the empty states of bulk VSe$_2$[23], which could contribute to the dI/dV spectroscopy. Shown as the blue arrow in Figure 3f, the dI/dV image of empty states presents stripe structures parallel to the Se dimerization direction. From the line profile analysis, the stripes of dI/dV image are laterally shifted by ~ 0.6 Å from those of topographic image (Figure S8). In addition, similar modulations are observed along the stripes in both Figure 3e and 3f (Figure S8g). These results suggest that the V atoms are accordingly dimerized along the same direction of Se dimerization.

In addition to the MIT at 135 K, we note that the partial gap persists up to $T_{CDW} = 350 \pm 8$ K due to the perfect FS nesting (Figure 2g). In order to confirm the existence of the CDW ordering in our STM measurements, we further analyze the 2D FFT images in Figures 3b and 3d. The two FFT images contain similar peaks, marked by white circles. Selectively filtered STM images, highlighting the pink and white circles, reveal the presence of a weak CDW order at both 79 K (Figure 3h) and 300 K (Figure 3i). The periodicity of the CDW is measured to be 4a (a = lattice constant) in the line profiles of Figure 3j, which compares well with the unidirectional CDW order in the bulk.[11]

Based on our ARPES and STM measurements, we propose a schematic phase diagram for the electronic reconstruction of VSe$_2$ systems in the parameter plane describing film thickness and temperature (Figure 4a). When the sample dimensionality is reduced from 3D to 2D, the weakly nested FS (phase (i)) is transformed into the perfectly nested FS with elongated parallel sides (phase (ii)), resulting in a significant increase of $T_{CDW}$. Lowering the temperature induces partial gap opening in the nested sections (red dashed lines) of the FS for both the 3D (phase (iii)) and 2D (phase (iv)) CDW phases, while their residual parts (blue solid lines) remain ungapped. The further introduction of interfacial effects in the 2D heterostructure, such as a lattice mismatch, opens a complete gap in the FS (phase (v), red and blue dashed lines), indicative of an MIT.

In order to show the universal RG flow toward perfect FS nesting in ML VSe$_2$, we construct an effective field theory, which consists of the following three main physical ingredients: (I) Hot electrons with imperfect FS nesting, (II) critical CDW fluctuations and (III) an effective Yukawa-type interaction between (I) and (II). All of the details were presented in our theoretical study.[24] Our theoretical analysis is parameterized in terms of the "tangential" velocity $v$ in the hot FS, which goes to zero when the hot FS becomes straight; the group velocity $c$ of CDW fluctuations, which is associated with the Q vectors in Figure 2b; and the Yukawa coupling constant $e$ between hot electrons and CDW order parameters. Considering a 3D system with imperfect FS nesting, one can show that $v$ decreases much more slowly than in the 2D case. Assuming the 3D dispersion $E = k_x + v k_y + k_z^n$ with n > 1, we find that relatively small fractions of the FS are coupled with CDW fluctuations and $e$ becomes irrelevant, which explains why the mean-field theory works well in 3D.[25-29] On the other hand, strong CDW fluctuations in 2D give rise to self-consistent renormalization effects for both hot electrons and CDW excitations, which cause $v$ to vanish. Figure 4b shows that the RG flow of these three parameters approaches zero in 2D. Even if $e$ flows to zero, this does not mean that the electrons and CDW order parameters are decoupled. They are still strongly interacting, where the ratio of $e^2/v$ flows into a finite fixed point value.[24] This result indicates that reduced dimensionality to 2D leads to strong CDW order and causes perfect FS nesting, consistent with the above experimental observations.

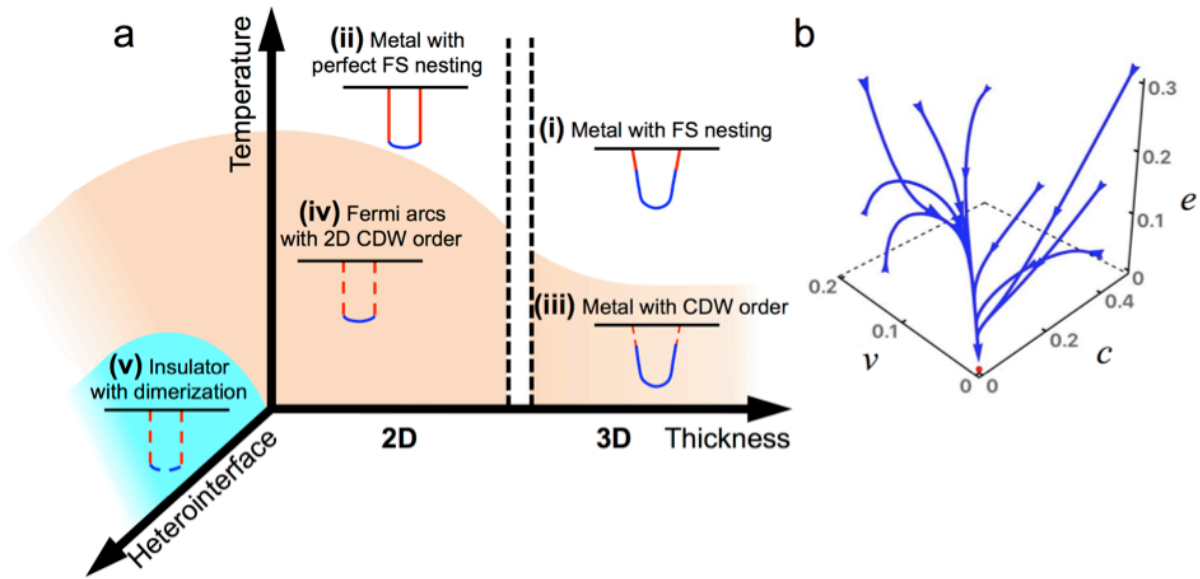

**Figure 4.** Schematic phase diagram for the electronic reconstruction of VSe$_2$. (a) Schematic phase diagram in parameter space defined by thickness, lattice mismatch, and temperature, based on both ARPES and STM measurements. Each phase contains the schematic FS model, where solid (dashed) contours correspond to the ungapped (gapped) section in one-sixth of the Brillouin zone. This phase diagram summarizes how strongly correlated electron physics in ML VSe$_2$ emerges from that of weakly interacting electrons in 3D bulk VSe$_2$, both by reducing the film thickness to the ML limit and by introducing a heterointerface with graphene. (b) The renormalization group flow diagram for the fermion-boson interaction parameter ($e$) and both electron and order parameter velocities ($v$ & $c$). All these parameters flow in a zero fixed-point value at low temperature, which confirms the emergence of perfect FS nesting in 2D.

Recently, room-temperature ferromagnetism has been reported in ML VSe$_2$ from superconducting quantum interference device (SQUID) measurements.[6] We also performed SQUID measurements on our ML VSe$_2$ and detected a magnetic hysteresis signal (Figure S9). The observed magnetic signature does not show any correlations with our two types of charge ordering transitions and full energy gap at low temperature (Figure S10). In principle, it is quite difficult to imagine how Stoner ferromagnetism arises in our CDW or insulating states. Disorder may play a role in the emergence of localized magnetic moments, which deserves to be investigated further.

The dimensionality and interfacial interactions of quantum materials are essential factors for controlling the many-body interactions responsible for electronic reconstruction phenomena, which will have profound implications for device applications. The present study demonstrated two distinct electronic reconstructions in a van der Waals heterostructure based on ML VSe$_2$ and graphene. Using ARPES, STM, and RG theory, we found that the quasiparticle dynamics of hot electrons becomes effectively one dimensional in ML VSe$_2$, causing a dramatic enhancement of the CDW ordering temperature. Second, the interfacial coupling between ML VSe$_2$ and graphene such as lattice mismatch, misalignment, charge transfer, and hybridization, triggers the emergence of a novel MIT phase.

**METHODS**

**Molecular beam epitaxy.** ML VSe$_2$ was grown on epitaxial bilayer graphene (BLG) on SiC using a home-built molecular beam epitaxy (MBE) ultra-high vacuum (UHV) system with a base pressure of 2 × 10$^{-10}$ Torr. We used 6H-SiC(0001) substrates, supplied by the Crystal Bank at Pusan National University. The SiC substrates were outgassed at 650°C for a few hours and then annealed three times up to 1300°C for 2 min to form BLG, as verified by reflection high-energy electron diffraction and low-energy electron diffraction[30] (Figure S2). High-purity V (99.8%) and Se (99.999%) were simultaneously evaporated by an electron-beam evaporator and a Knudsen cell, respectively, onto a substrate maintained at 250°C. The growth process was monitored with in situ RHEED, and the growth rate was 5 min per monolayer. After growth, the sample was annealed at 450°C for 30 min and then capped with 100 nm of Se by deposition at room temperature. After transferring the sample through air to the UHV chambers for ARPES or STM, the Se capping layer was evaporated by annealing the sample to 300°C for several hours in UHV before the subsequent characterization.

**Angle-resolved photoemission spectroscopy measurements.** ARPES measurements were performed in the microARPES end-station (base pressure ~3 × 10$^{-11}$ torr) at the MAESTRO facility at beamline 7.0.2 at the Advanced Light Source, Lawrence Berkeley National

Laboratory. The ARPES system was equipped with a VG Scienta R4000 electron analyzer. The lateral size of the synchrotron beam was estimated to be between 30 and 50 μm. We used photon energies of 105 eV to map the $VSe_2$ Brillouin zone and 48 eV and 105 eV for fine temperature dependent scans. The total energy resolution was better than 20 meV, and the angular resolution was 0.1°. For band mapping along $k_z$, a series of measurements was carried out with various photon energies in the range of 65-160 eV.[10, 31]

**Scanning tunneling microscopy and spectroscopy measurements**. STM and STS measurements were carried out using a home-built low-temperature STM system (base pressure ~7 × $10^{-11}$ Torr).[32] Tungsten tips were prepared via electrochemical etching and cleaned with electron beam heating. The bias voltages stated in the topographic images were applied to the sample. The dI/dV mapping was performed with the standard lock-in technique, with a modulation signal of 10 mV at the frequency of 1.2 kHz. All STM and STS measurements were carried out at 79 K, except for a few room temperature data, as noted separately.

**Renormalization group analysis.** It turns out that perturbative RG analysis cannot be applied to the present problem in a straightforward manner. When there are FSs, these FS electrons are strongly correlated near quantum phase transitions in 2D, referred to as FS problems.[33] Recently, the technique of "graphenization" has been proposed as a way to controllably evaluate Feynman diagrams in the FS problem. This methodology generalizes the dimensional regularization technique for interacting boson problems into the FS problem, where the density of states is reduced to allow control of effective interactions of electrons.[34] Based on this recently developed dimensional regularization technique, we perform the perturbative RG analysis, in which all interaction parameters are renormalized self-consistently.

■ ASSOCIATED CONTENT

Supporting Information

Additional information on the structural and chemical characterization of VSe$_2$, ARPES data of bulk VSe$_2$ and temperature-dependence of ML VSe$_2$, detailed analysis of 2D FFT images and dI/dV data of the insulating phase at 79 K, and magnetic measurements of 1.5 ML VSe$_2$.

■ AUTHOR INFORMATION

**Corresponding Authors**

* E-mail: tkfkd@postech.ac.kr.

* E-mail: kimjd@ulsan.ac.kr.

* E-mail: yjchang@uos.ac.kr.

**Author Contributions**

G.D., B.C., and I.J. contributed equally to this work.

**Notes**

The authors declare no competing financial interest.

■ ACKNOWLEDGMENTS

This work was supported by National Research Foundation (NRF) grants funded by the Korean government (No. NRF-2009-0093818, NRF-2014R1A4A1071686, NRF-2015R1D1A1A01057271, and NRF-2017R1C1B2004927). S.U. acknowledges financial support from the Danish Council for Independent Research, Natural Sciences under the Sapere Aude program (Grant No. DFF-4090-00125) and from VILLUM FONDEN (grant no. 15375). The Advanced Light Source is supported by the Director, Office of Science, Office of Basic Energy Sciences, of the U.S. Department of Energy under Contract No. DE-AC02- 05CH11231. K.-S.K. was supported by the Ministry of Education, Science, and Technology (No. NRF-2015R1C1A1A01051629 and No. 2011-0030046) of the National Research Foundation of Korea (NRF) and by the TJ Park Science Fellowship of the POSCO TJ Park Foundation, and by the POSTECH Basic Science Research Institute Grant (2017). K.-S.K appreciated the great hospitality of the Asia Pacific Center for Theoretical Physics (APCTP). Work at IBS CCES was supported by the Institute for Basic Science in Korea (Grant No. IBS-R009-G1). The authors thank Y.-H. Chan, M.Y. Chou, S.-H. Lee, B.L. Chittari, and J. Oh for fruitful discussion.